\documentclass[apj]{emulateapj}
\usepackage{subfigure}
\usepackage{times}
\usepackage{graphicx}
\usepackage{amssymb}
\usepackage{multirow}
\usepackage{color}
\usepackage{wrapfig}
\usepackage{float}
\usepackage[flushleft]{threeparttable}
\usepackage{lineno}
\usepackage{amsmath}
\usepackage{url}

\newif\ifAMStwofonts
\AMStwofontstrue

\makeatletter

\newcommand{\Rmnum}[1]{\expandafter\@slowromancap\romannumeral #1@}
\makeatother

\shorttitle{ Magnetic field decay of PSR J1640$-$4631}

\shortauthors{Gao et al. 2017}

\begin{document}

\title{The Dipole Magnetic Field  and Spin-down Evolutions of The High
          Braking Index Pulsar PSR J1640$-$4631}
\author{Zhi-Fu Gao\altaffilmark{1,2},~
Na Wang\altaffilmark{1,2}, Hao Shan\altaffilmark{1}, Xiang-Dong Li \altaffilmark{3},Wei Wang~\altaffilmark{4,5,\dag}~}

\altaffiltext{1}{Xinjiang Astronomical Observatory, Chinese Academy of Sciences, 150, Science 1-Street, Urumqi, Xinjiang, 830011, China; zhifugao@xao.ac.cn}
\altaffiltext{2}{Key Laboratory of Radio Astronomy, Chinese Academy of Sciences, West Beijing Road, Nanjing, Jiangsu 210008, China }
\altaffiltext{3}{Department of Astronomy and Key Laboratory of Modern Astronomy and Astrophysics, Nanjing University, Jiangsu 210046, China}
\altaffiltext{4}{School of Physics and Technology, Wuhan University, Wuhan, Hubei, 430072, China; wangwei2017@whu.edu.cn}
\altaffiltext{5}{National Astronomical Observatories, Chiese Academy of Sciences, 20A Datun Road, Chaoyang District, Beijing 100012, China;wangwei2017@whu.ac.cn}
\begin{abstract}
In this work, we interpreted the high braking index of PSR J1640$-$4631 with a combination of the magneto-dipole radiation and dipole magnetic field decay models. By introducing a mean rotation energy conversion coefficient $\overline{\zeta}$, the ratio of the total high-energy photon energy to the total rotation energy loss in the whole life of the pulsar, and combining the pulsar's high-energy and timing observations with reliable nuclear equation of state, we estimate the pulsar's initial spin period, $P_{0}\sim (17-44)$\,ms, corresponding to the moment of inertia $I\sim (0.8-2.1)\times 10^{45}$\,g\,\,cm$^{2}$.  Assuming that PSR J1640$-$4631 has experienced a long-term exponential decay of the dipole magnetic field, we calculate the true age $t_{\rm age}$, the effective magnetic field decay timescale $\tau_{\rm D}$, and the initial surface dipole magnetic field at the pole $B_{p}(0)$ of the pulsar to be $2900-3100$\,yr, $1.07(2)\times10^{5}$\,yr, and $(1.84-4.20)\times10^{13}$\,G, respectively. The measured braking index of $n=3.15(3)$\, for PSR J1640$-$4631 is attributed to its long-term dipole magnetic field decay and a low magnetic field decay rate, $dB_{\rm p}/dt\sim -(1.66-3.85)\times10^{8}$ G yr$^{-1}$. Our model can be applied to both the high braking index ($n>3$) and low braking index ($n<3$) pulsars, tested by the future polarization, timing, and high-energy observations of PSR J1640$-$4631.
\end{abstract}
\keywords{stars: evolution -- magnetic field: neutron --  pulsars: individual (J1640$-$4631) -- supernova remnant: general}

\section{Introduction}
Pulsars are commonly recognized as highly magnetized and
rapidly rotating neutron stars (NSs).
 A pulsar's secular spin-down is mainly caused by its rotational energy losses (Lyne et al. 2015). An important and measurable quantity closely related to a pulsar's  rotational evolution is the braking index $n$, defined by assuming that the star spins down in the light of a power law\,(Lyne et al. 1993)
\begin{equation}
  \dot{\Omega}=~-K \Omega^{n}~,
  \end{equation}
where $\Omega$ and $\dot{\Omega}$ are the angular velocity and its derivative of the star, respectively, and $K$ is a proportionality constant.

A standard way to define the braking index is
\begin{equation}
  n\equiv ~\frac{\nu\ddot{\nu}}{\dot{\nu}^{2}}= 2-\frac{P\ddot{P}}{\dot{P}^{2}},
\end{equation}
where $\ddot{\Omega}$ is the second derivative of $\Omega$,
$\nu=\Omega/2\pi$ the spin frequency, and $P=1/\nu$ the spin
period (Manchester \& Taylor 1977, and references therein).
When the magneto-dipole radiation (MDR) solely causes the pulsar to spin down, the braking index is predicted to be $n=3$.
Since the braking index of a pulsar can provide some information about the pulsar's energy loss mechanisms, it has been investigated by many authors (for recent works, see, e.g., Espinoza et al. 2011, Ferdman et al. 2015; Franzon et al. 2016; Rogers \& Safi-Harb 2017).

Until now, only 9 of the $\sim$ 2500 known pulsars\footnote{see the ATNF catalogue (Manchester et al. 2005)} have  reliably measured braking indices. Most of them are remarkably lower than 3, demonstrating that the spin-down mechanism is not pure MDR. Recently, Dupays et al.\,(2008, 2012) put forward an energy loss mechanism, called quantum vacuum friction\,(QVF), which results from the
interaction between the magnetic dipole moment of a pulsar and its induced quantum vacuum. Taking into account the QVF effect, the measured
 braking indices of $n<3$ can be well explained\,(Coelho et al. 2016). To account for the observed braking indices, several other interpretations have been put forward\,(see Gao et al. 2016 for a brief summary).

Recently, Gotthelf et\,al.\,(2014) reported the discovery
of PSR J1640$-$4631, associated with the TeV $\gamma-$ray source HESS J1640$-$465 using the $NuSTAR$ X-ray observatory. Based on its timing observational data, $\nu=4.84341$\,s$^{-1}$, $\dot{\nu}=-2.2808\times10^{-11}$\,s$^{-2}$ and $\ddot{\nu}=3.38(3)\times10^{-22}$\,s$^{-3}$,\,(corresponding to $P=206.4$ ms, $\dot{P}=9.7228\times10^{-13}$\,s~s$^{-1}$ and $\ddot{P}=-5.27(13)\times10^{-24}$\,s~s$^{-2}$, respectively), Archibald et\,al.\,(2016) derived its braking index to be $n=3.15(3)$\,(here and following all digits in parentheses denote the standard uncertainty). PSR J1640$-$4631 is the first pulsar with a measured braking index larger than 3.

The possible origin of the braking index of PSR J1640$-$4631 has been discussed in the literature. The decrease in the magnetic inclination angle $\alpha$\,(the angle between the rotational and the magnetic axes) can result in a braking index higher than 3\,(Ek\c{s}i et al. 2016). A combination of gravitational wave\,(GW) and magnetic energy dipole loss mechanism radiation could give rise to a braking index $3<n<5$\,\,(de Araujo et al. 2016 a,\,b). Alternatively, such a braking index can be accounted for if the magnetic dipole braking still dominates but the star is experiencing a long-term dipole magnetic field decay.

The surface dipole magnetic field has a significant effect on the spin evolution of a pulsar, as well as on its braking index\,(e.g., Goldreich \& Reisenegger 1992; Tauris \& Konar 2001; Jones 2009; Marchant et~al. 2014). In the case of a varying dipole magnetic field at the pole $B_{\rm p}$, the braking index $n$ can be simply expressed as follows:
\begin{equation}
n=3-4\frac{\dot{B}_{\rm p}}{B_{\rm p}}\tau_{\rm c}\equiv 3-4\frac{\tau_{\rm c}}{\tau_{\rm B}},
\end{equation}
where $\tau_{\rm c}=P/2\dot{P}$ is the characteristic age and $\tau_{\rm B}=B_{\rm p}/\dot{B_{\rm p}}$ is the timescale on which the dipole component evolves. The above equation shows that any variation in $B$ can lead to a deviation from 3. For a decreasing $B_{\rm p}$, we will always obtain $n>3$ owing to a decaying dipole braking torque\,(e.g., Rea \& Esposito 2011; Kaminker et al. 2014; Potekhin et al. 2015), whereas $n<3$ for an increasing $B_{\rm p}$\,(e.g., Mereghetti 2008; Gao et al. 2016).

The motivation of this work is to account for the high $n$ of PSR J1640$-$4631 within a theoretical model of a long-term dipole magnetic field decay. The rotational evolution of PSR J1640$-$4631 is assumed to be dominated by the magnetic field anchored to the solid crust of the star. Since the core field evolves on much longer timescales, the crustal field evolves mainly through ohmic dissipation and the
Hall drift. The Hall drift mainly occurs in the neutron-drip solid or out of the core, conserving the magnetic energy\,(e.g., Haensel et al. 1990; Hertman et al. 1997; Page et al. 2000), while ohmic diffusion occurs in the range of density below the neutron-drip threshold and dissipates magnetic energy through Joule heating\,(e.g., Pons et al. 2009, 2012; Gourgouliatos \& Cumming 2015).

The remainder of this work is organized as follows: The key problems in estimating the true age of PSR J1640$-$4631 are presented in Section 2, and the initial spin period is constrained in Section 3, The initial magnetic field strength, the true age, and the effective dipole field evolution timescale of the pulsar are determined in Section 4, the dipole magnetic field and spin-down evolutions are numerically simulated in Section 5, and alternative dipole magnetic field decay models are presented in Section 6. Comparisons with other works and discussions are given in Section 7.
\section{Key problems in estimating $t_{\rm age}$}
To investigate the long-term spin evolution of a pulsar, it is best to know the true age of the pulsar. It is well known that $\tau_{\rm c}$ is a poor age approximation, and it coincides with the true age only if the current spin period is much longer than the initial value and the braking torque and the moment of inertia $I$ are constant during the entire pulsar life. A pulsar's true age $t_{\rm age}$ can be estimated by the age of its associated supernova remanant (SNR), because it is universally considered that pulsars originate from supernova explosions\,(e.g., Kouveliotou et al. 1994; Gaensler et al. 2001; Vink \& Kuiper 2006).

The SNR ages are currently estimated based on the measurements of the shock velocities, the X-ray temperatures, and/or other quantities. Assuming that an SNR is in the Sedov phase\footnote{Sedov (1959) showed that there is a self-similar solution for
the adiabatic expansion phase, and firstly applied this solution
to estimate the age of SNRs in this phase, i.e., the Sedov age. The Sedov age estimates rely on many assumptions, but ultimately on the SNR size and the SNR temperature. If a SNR is too young (when the expansion is free), or too old (when radiation energy losses brake the expansion), the Sedov expansion is no longer applicable.}\,(Sedov 1959), the SNR age can be estimated by a convenient expression:
\begin{equation}
t_{\rm SNR}\approx 435\times R_{\rm SNR}T^{-1/2}~\rm~yrs,
\end{equation}
where the SNR radius $R_{\rm SNR}$ is in units of pc, and the post-shock plasma temperature $T$ is in units of keV (for details see Gao et al. 2016).

G338.3$-$0.0 is a shell-type, 8$^{'}$ diameter SNR\,(Shaver \& Goss 1970; Whiteoak \& Green 1996), and spatially relates with HESS J1640$-$465, which is considered to be the the most luminous $\gamma$-ray source in the Galaxy. The X-ray pulsar PSR J1640$-$4631 was recently discovered within the shell of SNR G338.3$-$0.0 \,(Gotthelf et al. 2014). Unfortunately, no X-ray emission was detected from the shell of the SNR\,(probably due to its large intervening column density, $N_{\rm H} \sim1.4\times10^{23}$\,cm$^{-2}$ and low temperature; Lemiere et al. 2009; Castelletti et al. 2011); thus, the true age of PSR J1640$-$4631 cannot be estimated from Equation (4).

The pulsar's true age can be inferred from Equation (1),
\begin{eqnarray}
&&t=\frac{P}{(n-1)\dot{P}}~[1- (\frac{P_{0}}{P})^{n- 1}]~~~~\left(n\neq 1\right),\nonumber\\
&&t=2\tau_{\rm c}~\rm{\ln} (\frac{P}{P_{0}})~~~~~~~~~~~~~~~~~\left(n= 1\right),
\end{eqnarray}
where $P_{0}$ is the initial spin period of the star and the braking index is assumed to be a constant since its birth\,(e.g., Manchester et al. 1985; Espinoza et al. 2011; Rogers \& Safi-Harb 2017). As discussed in Gao et al.\,(2016), one indeed cannot derive an analytical expression such as Equation (5) to estimate $t_{\rm age}$ if we consider that either $K$ or $n$ in Equation (1) is a time-dependent quantity.  Since the braking index is influenced by the variation of braking torque during the observational intervals, it is necessary to assume a constant mean braking index to calculate the age of a young pulsar by using the above equation (Gao et al. 2016).

The mean braking index $\overline{n}$ of a pulsar is defined as
\begin{equation}
\overline{n}=\int n(t)dt/\int dt.
\end{equation}
As seen from Equation (5), $t_{\rm age}$ is less than $\tau_{\rm c}$ if $\overline{n}=3$. In addition, Equation (5) implies an upper limit of the true age, $t_{\rm age}=2\tau_{\rm c}/(\overline{n}-1)$, corresponding to $(P_{0}/P)^{\overline{n}-1}=0$. However, it is difficult to obtain the values of $P_{0}$ and $\overline{n}$ of the pulsar; $t_{\rm age}$ cannot be directly derived from Equation (5).
\section{Initial spin period of PSR J1640$-$4631 }
The initial spin period $P_{0}$ of a pulsar is also very important in our ability to investigate the pulsar's long-term spin evolution.
Since one cannot know the value of $P_{0}$ of PSR J1640$-$4631,
theoretically estimating its initial spin period becomes urgent. This section is composed of the following two subsections.
\subsection{X-ray and $\gamma$-ray luminosities}
As mentioned above, PSR J1640$-$4631 is located inside G338.1$-$0.0, the pulsar wind nebula\,(PWN), which is primarily
responsible for the $\gamma$-ray emission of HESS J1640$-$465.
The detection of the pulsar in the SNR provides long-awaited evidence that a PWN powers the $\gamma$-ray source HESS J1640$-$465, and its properties can be used to test the radiation mechanisms\,(e.g., Reynolds 2008; Slane et al. 2014).

Recently, employing Markov Chain Monte Carlo algorithm, Gotthelf et al. (2014) fitted the parameters of the PWN in G338.3$-$0.0 and obtained the best-fitted values of free parameters, including the high-energy
photon (``photon '' in short) fluxes. For the purpose of illustration, in Table 1 we list the photon fluxes of PSR J1634$-$4731 and its PWN, respectively.
\begin{table}
\caption{ Spectrum Fluxes of PSR J1640$-$4631 and Its Wind Nebula.  }
\begin{tabular}{cccc}\\
\hline 
Flux & Observed Flux & Telescope & Reference \\
 \hline
$F_{\rm X}$[2-25\,keV] &$1.0\times10^{-12}$   & $NuSTAR$ & Gotthelf et~at. (2014)\\
$F_{\rm HESS}$[10-500\,GeV] &$1.6\times10^{-11}$   & H.E.S.S &Gotthelf et~at. (2014)\\
\hline\\
\end{tabular}\\
\label{tb:1}
\end{table}

In Table 1, the spectrum flux of $F_{\rm X}$[$2-25$\,keV]
includes the X-ray fluxes of PSR J1634$-$4731 and its PWN. All the given fluxes are in units of $\rm erg\, cm^{-2}s^{-1}$.
Then, the total luminosity (the currently measured photon luminosity) is calculated as
\begin{equation}
  L_{X,\gamma}=(F_{\rm PSR}+F_{\rm PWN})4\pi d^2=2.93\times10^{35}d_{12}^{2}\,{\rm erg\,s^{-1}},
\end{equation}
 where $d_{12}$ is a dimensionless distance in units of 12\,kpc.
 \subsection{Mean rotational energy transformation coefficient}
In order to estimate $P_{0}$ of
PSR J1634$-$4731, we introduce an energy transformation
coefficient, $\varsigma$, which is the ratio of the total photon luminosity, $L_{X,\gamma}$, to the spin-down luminosity of the pulsar $L_{\rm spin}$\,($L_{\rm spin}=I\dot{\Omega}(t)\Omega(t)$)
\begin{equation}
\zeta=\frac{L_{X,\gamma}}{L_{\rm spin}}=\frac{(F_{\rm PSR}+F_{\rm PWN})4\pi d^2}{4\pi^{2}I\dot{P}P^{-3}}.
\end{equation}
From the values of current $P$ and $\dot{P}$, we get the current luminosity $L_{\rm spin} \sim 4.36\times10^{36}I_{45}$\,erg\,s$^{-1}$, where $I_{45}$ is the moment of inertia in units of $10^{45}\,\rm g\,cm^{2}$. Inserting the values of $L_{X,\gamma}$ and $L_{\rm spin}$ into Equation (8) yields a general solution for the current coefficient, $\zeta=0.067d_{12}^{2}I^{-1}_{45}$.  Recently, Getthelf et al. (2014) obtained the best-fit source distance to G338.3$-$0.0, $d\sim$12\,kpc. This distance agrees well with the typical supernova explosion scenario\,(e.g., the order of magnitude of explosion energy is $10^{51}$ erg). In this paper, we adopt the distance of $d_{12}=1$.

Since $\zeta$ is variable with time, in order to estimate the initial spin parameters\,including $P_{0}$, a constant $\zeta$ should be assumed in a specific model. We define a mean rotation energy transformation coefficient $\overline{\zeta}$,
 \begin{equation}
 \overline{\zeta}=\frac{E_{X,\gamma}}{E_{\rm spin}}=
 \frac{\int^{t}_{0}L_{X,\gamma}(t)dt}{E_{\rm spin}},
 \end{equation}
where $E_{X,\gamma}$ is the total photon energy and $E_{\rm spin}$ is the total rotation energy loss, which can be estimated by
\begin{equation}
 E_{\rm spin}=
 \int^{t}_{0}L_{\rm spin}(t^{'})dt^{'}\simeq \frac{2{\pi}^{2}I}{P_{0}^{2}},
 \end{equation}
for $P_{0}\ll P$ (Tanaka Shuta 2016). For a constant braking index $\overline{n}$, the spin-down luminosity $L_{\rm spin}(t)$
evolves in the light of a simple power-law form,
\begin{equation}
L_{\rm spin}(t)=L_{\rm spin}(0)\left(1+\frac{t}{\tau_0}\right)^{-\frac{\overline{n}+1}
{\overline{n}-1}},
\end{equation}
where $L_{\rm spin}(0)$ is the initial spin-down
luminosity and $\tau_{0}$ is the initial spin-down timescale of the star, which is defined as
\begin{equation}
\tau_0=\frac{P_0}{(\overline{n}-1)\dot{P_0}}=\frac{2\tau_{\rm c}}{\overline{n}-1}
\left(\frac{P_{0}}{P}\right)^{\overline{n}-1}.
\end{equation}
with the initial period derivative $\dot{P_0}$. Since PSR J1640$-$4631 is a pulsar powered by rotational energy loss, the total photon luminosity, $L_{X,\gamma}(t)$, as well as $L_{\rm spin}(t)$, should decrease with time. For the sake of simplicity, we assume that $L_{X,\gamma}$ and $L_{\rm spin}(t)$ have the same evolution form.
Then the total photon energy from $t=0$ to $t=t_{\rm age}$ is calculated as
\begin{eqnarray}
&&E_{X,\gamma}
=\int_{0}^{t_{\rm age}}L_{X,\gamma}(0)(1+
\frac{t}{\tau_{0}})^{-(\frac{\overline{n}+1}{\overline{n}-1})}dt\nonumber\\
&&=L_{X,\gamma}(t_{\rm age})\cdot\left[(\frac{P_0}{P})^{-2}
-1\right]\cdot\tau_{\rm c}.
\end{eqnarray}
where $L_{X,\gamma}(0)$ is the initial total photon luminosity and the relations of $\tau_{0}+t_{\rm age}=\frac{2\tau_{\rm
c}}{\overline{n}-1}$ and $\tau_{0}=\frac{2\tau_{\rm c}}{\overline{n}-1}
(\frac{P_{0}}{P})^{\overline{n}-1}$ are utilized.

Inserting Equations (10) and (13) and $\tau_{c}\simeq 3360$\,yrs into Eq.(9), we obtain the initial spin period
\begin{equation}
P_{0}\simeq 20.8(\overline{\zeta}/0.067)^{-1}d_{12}^{-2}I_{45} ~~{\rm ms}.
\end{equation}
It is interesting to note that, although $\overline{n}$ has been used in the derivation above, the ultimate value of $P_{0}$ is free of $\overline{n}$. Recently, Abdo et al.\,(2010, 2013) presented catalogs of high-energy gamma-ray pulsars detected by the Large Area Telescope on the $Fermi$ satellite and estimated the values of $\zeta$ for some sources, (here $\zeta$ is the ratio of the gamma-ray luminosity to rotational energy loss rate of a pulsar). According to Abdo et al.\,(2010, 2013), the range of $\zeta$ is about $(0.01-1.0)I^{-1}_{45}$, but its average value is about $0.08I^{-1}_{45}$, close to our estimate of $\zeta=0.067d^{2}_{12}I^{-1}_{45}$. In Equation (14), we have adopted fiducial NS parameters:\,mass $M=1.4\,M_{\bigodot}$, radius $R$=10 km, and the moment of inertia $I_{45}=1$. Strictly speaking, the true value of $\overline{\zeta}$ is unknown and may be constrained by specific NS nuclear equation of state (EOS).

From Equation (14), it is clear that $P_{0}$ is a function of moment of inertia $I$. In order to investigate how different values of $I$\,(or the mass [$M$]-equatorial radius [$R$] relation) modify $P_{0}$, we need to take into account the $M-R$ relation of the uniformly rotating NS and feasible EOSs.
\subsection{Nuclear EOS and moment of inertia}
 Very recently, to investigate the possibility that some of soft gamma-ray repeaters (SGRs) and anomalous X-ray pulsars (AXPs) family could be canonical rotation-powered pulsars, Coelho et al. (2017) adopted realistic NS structure parameters\,(e.g., mass, radius, and moment of inertia) instead of fiducial values and considered the issue related to the high-energy luminosity and the spin-down luminosity of SGRs and AXPs in detail. Their results provide useful constraints on the initial spin period $P_{0}$ of PSR J1640$-$4631.

As we know, the maximum NS mass predicted by EoSs is model dependent\,(e.g., Dutra et al. 2014; Li et al. 2016; Xia et al. 2016; Xia \& Zhou 2017; Zhou et al. 2017). The largest sample of measured NS masses available for analysis is publicly accessible online at http://www.stellarcollapse.org/, and has been updated by J. Lattimer. A similar analysis can be found in Valentim  et al. (2011), from which one can get a range of about $1-2$\,$M_{\bigodot}$ for the observational NS masses. The minimum NS mass, as well as the maximum NS mass, is still a matter of debate. Though the NS masses could be could be less than $1.0\,M_{\bigodot}$ from the viewpoint of the EOS, it is difficult to explain their formations from the supernova explosion mechanism. Pulsars having mass of $\sim 2.0\,M_{\bigodot}$ are confirmed by Demorest et al. (2010) and Antoniadis et al. (2013).
The typical nuclear matter EOSs obtained from \"{O}zel \& Freire (2016) predict that pulsars have maximum mass larger than $2.0\,M_{\bigodot}$,\, which prefers the APR (Akmal et al. 1998) model and the RMF (relativistic mean-field) model\,(e.g., Lalazissis et al. 1997; Toki et al. 1995). In the APR model, properties of dense nucleon matter and the structure of  NSs are studied using variational chain summation methods and the new Argonne v18 two-nucleon interaction (Av18). However, in this work, we adopt the APR3 model (one version of APR EOS) instead of the RMF model. The main reasons are as follows:
\begin{enumerate}
\item The RMF theory is based on effective coupling constants that take the correction effect into consideration. However, these coupling constants are density dependent, and a microscopic theory is needed to calculate them.
\item Akmal et al. (1998) investigated the properties of dense nucleon matter and the structure of NSs and provided an excellent fit to all of the nucleon-nucleon scattering data in the Nijmegen database. The authors not only considered the nonrelativistic calculations with Av18 and Av18+UIX (Urbana IX three-nucleon interaction) models for nuclear forces but also described the relativistic boost interaction model (denoted as $\delta v$) with and without three-nucleon interaction (UIX$^{*}$).
\item In this work, we prefer the APR3 model, e.g., Av18+$\delta v$+UIX$^{*}$ model, which provides a constraint on the maximum NS mass $M\leq 2.2\,M_{\bigodot}$. This model has also been preferred and included in recent works\,(e.g., \"{O}zel \& Freire 2016; Zhou et al. 2017), which limit the range of NS masses to $<2.2\,M_{\bigodot}$ because of the absence of any data to constrain the relation at higher mass.
\end{enumerate}

In order to calculate the moment of inertia $I$, it is of importance for us to adopt a better $M-R-I$ relation for NSs. In this work, we  prefer an improved approximation expression of $M-R-I$ with $M\geq 1\,M_{\bigodot}$, provided by Steiner et al.\,(2016),
\begin{eqnarray}
 \frac{I}{MR^{2}}&&\simeq 0.01+(1.200^{+0.006}_{-0.006})
\beta^{1/2}-0.1839\beta
\nonumber\\
&&-(3.735^{+0.095}_{-0.095})\beta^{3/2}+5.278\beta^{2}.
\end{eqnarray}
where $\beta=(\frac{M}{R}\frac{\rm{km}}{M_{\bigodot}})$. The above equation shows less uncertainty, especially for compactness typical of 1.4\,$M_{\bigodot}$ stars, and the smaller uncertainties result from an assumption that $M_{max}> 1.97 M_{\bigodot}$ (Steiner et al. 2016).
Combining Equation (15) with the APR3 EoS, we obtain $I_{45}\sim 0.81(1)-2.07(3)$, corresponding to $M\sim(1.0-2.2)\,M_{\bigodot}$, and $R\sim(11.6-10.3)$\,km, respectively\,($M$ decrease with increasing $R$). Figure 1 shows the relation of $M, R$ and $I$ for PSR J1640$-$4631 in the APR3 model.
\begin{figure}[bt]
\centering
\includegraphics[angle=0,scale=.88]{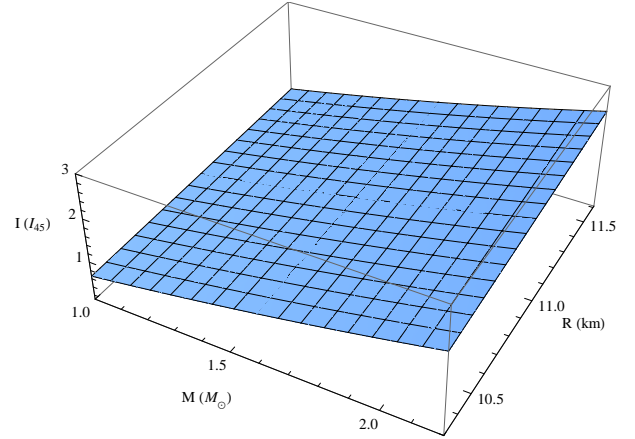}
\caption{Relation of $M, R$ and $I$ for PSR J1640$-$4631 in the APR3 model.}
\label{inertial}
\end{figure}
However, to effectively constrain the value of $P_{0}$, it is necessary to combine the EOS and the spin-down evolution and diploe magnetic field evolution equations. For details, see the next section.
\section{Constraining $B_{\rm p}(0)$, $t_{\rm age}$ and $\tau_{\rm D}$ of the pulsar }
\subsection{Hall drift and Ohmic decay}
Here we restrict ourselves to the dipole magnetic field evolution in the crusts of NSs, ignoring possible effects involving the fluid core. Assuming that ions are locked into a column lattice, the only freely moving charged species in the crust are electrons. In order to reconcile the measured braking index of PSR J1640$-$4631 with the dipole magnetic field decay, it is necessary to introduce the Hall induction equation describing the evolution of the crust magnetic field:
\begin{equation}
\frac{\partial \vec{B}}{\partial t}=-\nabla\times\left[\frac{c^{2}}{4\pi\sigma}\nabla\times (e^{\nu}\vec{B})
+\frac{c}{4\pi en_{e}}[\nabla\times(e^{\nu}\vec{B})]\times\vec{B}\right],
\end{equation}
where $\sigma$ is the electric conductivity parallel
to the magnetic field, $e^{\nu}$ is the relativistic
redshift correction, $n_{e}$ is the electron number density,
and $e$ the electron charge.
This equation contains two different effects that act on two distinct timescales, which can be estimated as
\begin{equation}
t_{\rm Hall}=\frac{4\pi n_{e}e L^2}{cB},~~~ t_{\rm Ohm}=\frac{4\pi \sigma L^{2}}{c^{2}},
\end{equation}
where $t_{\rm Ohm}$ is the ohmic dissipation timescale with a typical value of $\sim 10^{6}$ yr or more (Vigan\`{o} et al. 2013),
$t_{\rm Hall}$ is the Hall drift timescale, and $L$ is a characteristic length scale of variation, which can be taken to be the thickness of the neutron star crust\,(e.g., Haensel et al. 1990; Pons \& Gepport 2007; Pons et al. 2009; Aguilera et al. 2008; Ho 2011). There is a typical value of several $\times(10^{4}-10^{5})$\,yr for $t_{\rm Hall}$, which is constrained by the magnetic field strength of high magnetic field pulsars and magnetars (Vigan\`{o} et al. 2013).

 There are two field configurations: one in which the field is confined in the crust, and another in which the field extends into the core\,\,(there could be an electron current flowing through a superconducting core). For the purposes of illustration, we choose a dipole magnetic field $B_{\rm p}$ constrained in the crust and assume a simple exponential decay of $B_{\rm p}$ over time,
\begin{equation}
\frac{dB_{\rm p}}{dt}=-\frac{B_{\rm p}}{\tau_{\rm D}},
\end{equation}
as done in other works\,(e.g., Pons et al. 2009; Vigan\`{o} et al. 2013). This assumption requires that the pulsar has an initial magnetic field at the pole $B_{\rm p}(0)$\,(corresponding to $t=0$) and that the magnetic field decays at a rate proportional to its strength. $\tau_{\rm D}$ is an effective dipole magnetic field timescale, which is defined as
\begin{equation}
\frac{1}{\tau_{\rm D}}=\frac{1}{t_{\rm Hall}}+\frac{1}{t_{\rm Ohm}}.
\end{equation}
The Hall timescale $t_{\rm Hall}$ is universally one order of magnitude lower than the ohmic timescale (e.g., Pons \& Gepport 2007; Pons et al. 2009). This would imply that the effective timescale we define in Equation (19) is basically dominated by the Hall timescale.
By integrating Equation (18), we get a time-dependent field
\begin{equation}
B_{\rm p}(t)=B_{\rm p}(0)e^{-\frac{t}{\tau_{\rm D}}}.
\end{equation}
 With respect to two parameters of $\tau_{\rm D}$ and $B_{\rm p}(0)$, there are three aspects that need to be addressed explicitly:
\begin{enumerate}
\item In general, the effective timescales of pulsars are in the range of about $10^{4}-10^{7}$\,yrs\,(see Muslimov \& Page 1995, 1996; Geppert \& Rheinhardt 2002; Lyutikov et al. 2015;  Mereghetti et al. 2015 and references therein). However, $\tau_{\rm D}$ of PSR J1640$-$4631 is determined mainly by $t_{\rm Hall}$. To exactly constrain $\tau_{\rm D}$ of the star requires both observational and theoretical investigations.

 \item  Though there are several methods for roughly measuring
 the magnetic field strength of a pulsar, such as magneto-hydrodynamic pumping, Zeeman splitting, cyclotron lines, magnetar bursts and etc., a direct estimate of the strength $B_{\rm p}(0)$ from the observations is unavailable.
 \item For PSR J1640$-$4631, the initial dipole magnetic field was inherited from its progenitor star and cannot be estimated from its initial spin parameters via a simple MDR model. In this work, we have assumed a variable dipole magnetic field for the pulsar.
\end{enumerate}
\subsection{Equations of $P(t)$, $\dot{P}(t)$ and $\ddot{P}(t)$}
Assuming that the MDR of a pulsar with a decaying field
can be responsible for the high braking index of
PSR J1640$-$4631, we here constrain three parameters $B_{\rm p}(0)$, $t_{\rm age}$ and $\tau_{\rm D}$ of the pulsar.

If the magnetic field evolution of PSR J1640$-$4631 cannot be ignored and the dipole braking still dominates, according to Blandford \& Romani (1988), the braking law of the pulsar is reformulated as
\begin{equation}
\dot{\nu}(t)=-\frac{2\pi^{2}R^{6}}{3Ic^{3}}B^{2}_{\rm p}(t)\nu^{3},
\end{equation}
where a constant inclination angle $\alpha=90^{\circ}$  is assumed for the sake of simplicity. Integrating Equation (21) gives the spin frequency,
\begin{equation}
\nu^{-2}=\nu^{-2}_{0}+2\int^{t}_{0}\frac{2\pi^{2}R^{6}
}{3Ic^{3}}B^{2}_{\rm p}(t)dt^{'},
\end{equation}
where $\nu_{0}$ is the initial spin frequency of the pulsar. Then we get the spin period,
\begin{equation}
P(t)=\left[P^{2}_{0}+(\frac{2\pi^{2}R^{6}B^{2}_{\rm p}(0)}{3Ic^{3}})\cdot \tau_{\rm D}\cdot
(1-e^{\frac{-2t}{\tau_{\rm D}}})\right]^{1/2}.
\end{equation}
Utilizing the differential method, we get the first-order
derivative of the spin period $\dot{P}(t)$,
\begin{eqnarray}
&&\dot{P}(t)=(\frac{2\pi^{2}R^{6}B^{2}_{\rm p}(0)
}{3Ic^{3}})\times e^{\frac{-2t}{\tau_{\rm D}}}\times \nonumber\\
&& \left[P^{2}_{0}
+(\frac{2\pi^{2}R^{6}B^{2}_{\rm p}(0)
}{3Ic^{3}})\times \tau_{\rm D}\times
(1-e^{\frac{-2t}{\tau_{\rm D}}})\right]^{-1/2},
\end{eqnarray}
and the second-order derivative of the spin period $\ddot{P}(t)$,
\begin{eqnarray}
&&\ddot{P}(t)=-(\frac{2\pi^{2}R^{6}B^{2}_{\rm p}(0)
}{3Ic^{3}})^{2}\times e^{\frac{-4t}{\tau_{\rm D}}}\times  \nonumber\\
&& \left[P^{2}_{0}
+(\frac{2\pi^{2}R^{6}B^{2}_{\rm p}(0)
}{3Ic^{3}})\times \tau_{\rm D}\times
(1-e^{\frac{-2t}{\tau_{\rm D}}})\right]^{-3/2}\nonumber\\
&&-\frac{2}{\tau_{\rm D}}\times (\frac{2\pi^{2}R^{6}B^{2}_{\rm p}(0)
}{3Ic^{3}})\times e^{\frac{-2t}{\tau_{\rm D}}} \nonumber\\
&&\times\left[P^{2}_{0}
+(\frac{2\pi^{2}R^{6}B^{2}_{\rm p}(0)
}{3Ic^{3}})\times \tau_{\rm D}\times
(1-e^{\frac{-2t}{\tau_{\rm D}}})\right]^{-1/2}.
\end{eqnarray}
In the above three equations there are unknown variables of $B_{\rm p}(0)$, $t_{\rm age}$ and $\tau_{\rm D}$, which will be constrained by numerically simulating in the next subsection.
\subsection{The results of numerical simulations  }
If we combine Equations (14) and (15) with the APR3 EOS, take $t=t_{\rm age}$, and insert the current values of $P=206.4$\,ms $\dot{P}=9.7228\times10^{-13}$\,s\,s$^{-1}$ and  $\ddot{P}=-5.27(13)\times10^{-24}$\,s\,s$^{-2}$ into Equations (23)$-$ (25), then we obtain the values of $\tau_{\rm D}$, $t_{\rm age}$ and $B_{\rm p}(0)$ given a specific value of $P_{0}$\,(corresponding to a specific $I$ of the pulsar).

By numerically simulating, we make a plot of $B_{\rm p}(0)$ versus $I$ for PSR J1640$-$4631, as shown in Figure 2.
\begin{figure}[bt]
\centering
\includegraphics[angle=0,scale=.88]{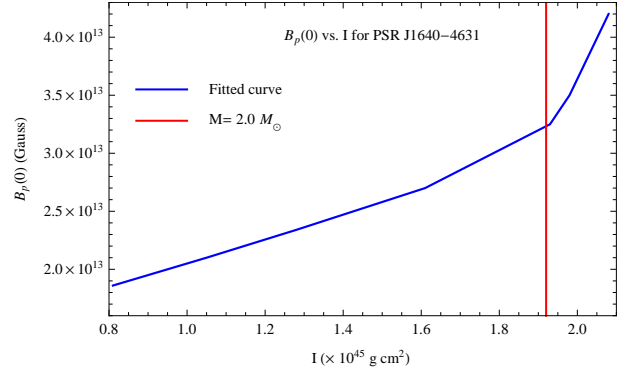}
\caption{Initial dipole magnetic field strength $B_{\rm p}(0)$ as a function of the moment of inertia $I$ for PSR J1640$-$4631. The range of $I$ is taken as $(0.80-2.09)I_{45}$, corresponding to $M\sim (1.0-2.2)\,M_{\bigodot}$.}
\label{index}
\end{figure}
 In Figure 2, the blue solid line denotes the fitted values of $B_{\rm p}(0)$ and $I$. The red solid line stands for a typical mass $M=2.0\,M_{\bigodot}$ ($I=1.92(5)$ and $R=10.94$ km) predicted by the APR3 EoS.  Also, we obtain the initial spin period $P_{0}\sim (17-44)$ ms, the true age $t_{\rm age}\sim 2800-3100$ yr, the initial dipole magnetic field strength $B_{\rm p}(0)\sim (1.84-4.20)\times10^{13}$ G, and the effective timescale $\tau_{\rm D}\sim 1.07(2)\times10^{5}$ yr for the pulsar.

As we know, the estimates of the crustal ohmic and
Hall timescales represent the choice of microscopic parameters.  Recently, Gourgouliatos \& Cumming (2014) estimate the average initial value of $\tau_{\rm Hall}$ for young pulsars about $100-200$ kyr. This timescale is close to that given by us.
\section{ Numerical Simulations of the Dipole Magnetic Field and Spin-down evolutions }
\subsection{The evolution of the braking index }
After obtaining $B_{\rm p}(0)$, $t_{\rm age}$ and $\tau_{\rm D}$ of the pulsar, we then use them in numerically simulating the spin-down evolution, especially accounting for its high braking index.

Combining Equations (23)-(25) with
$n=2-\frac{P\ddot{P}}{\dot{P}^{2}}$, we get
\begin{eqnarray}
&&n=3+\frac{3Ic^{3}}{\pi^{2}R^{6}B^{2}_{\rm p}(0)
e^{\frac{-2t}{\tau_{\rm D}}}\times\tau_{\rm D}}\nonumber\\
&& \times\left[P^{2}_{0}
+(\frac{2\pi^{2}R^{6}B^{2}_{\rm p}(0)
}{3Ic^{3}})\times\tau_{\rm D}\times
(1-e^{\frac{-2t}{\tau_{\rm D}}})\right].
\end{eqnarray}
From the above equation, it is easily seen that $n=3$ if $\tau_{\rm D}\rightarrow \infty$. In order to investigate the evolution of $n$, we plot the diagrams of $n$ versus $t$ for the pulsar.
\begin{figure}[bt]
\centering
\includegraphics[angle=0,scale=.88]{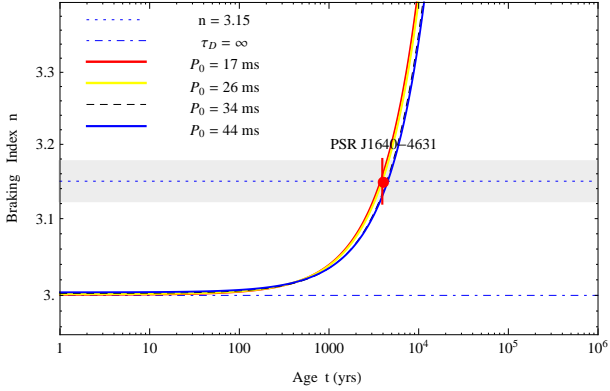}
\caption{Braking index as a function of $t$ for PSR J1640$-$4631.}
\label{index}
\end{figure}
In Figure 3, the red solid line, yellow solid line, black dashed line, and blue solid line stand for the predictions of the dipole magnetic field decay model with $P_{0}=17, 26, 34$ and 44\,ms, respectively. The blue dot-dashed line stands for the prediction of the MDR model, in which we assume the NS mass $M=1.4\,M_{\bigodot}$, radius $R=10$\,km and momentum of inertial, $I=10^{45}$\,g~cm$^{2}$, corresponding to $P_{0}\sim 21$ ms, $t_{\rm age}\sim 3220$ yr, $\tau_{\rm D}\sim 1.07\times10^{5}$\,yr and $B_{\rm p}(0)\sim 2.97\times10^{13}$\,G. Notice that the signs in Figure 3 are universally adopted in other figures in the subsequent sections of this work. The horizontal blue dotted line and the surrounding shaded region denote, respectively, the measured braking index of $n=3.15$ and its possible range given by the uncertainty 0.03 of PSR J1640$-$4631. From Figure 3, it is obvious that $n$ increases with $t$ owing to the decay of the dipole magnetic field.
\subsection{The evolution of the dipole magnetic field }
By using Eq.(20) and the constrained parameters of $t_{\rm age}$, $\tau_{\rm D}$ and $B_{\rm p}(0)$, we estimate the present value of the dipole magnetic field strength, $B_{\rm p}(t_{\rm age})\sim (1.77-4.10)\times10^{13}$\,G. This range is different from that estimated by the characteristic magnetic field at the polar of a pulsar $B_{\rm c}\simeq 6.4\times10^{19}\times(P\dot{P}I_{45})^{1/2}=2.87\times10^{13}$\,G, because the latter assumes a constant dipole braking torque and a constant momentum of inertial $I_{45}=1$. Then we obtain a mean magnetic field decay rate of the pulsar, $\Delta B_{\rm p}/\Delta t=(B_{\rm p}(t_{\rm age})-B_{\rm p}(0))/t_{\rm age}\approx -(2.41-3.51)\times10^{7}$\,G\,yr$^{-1}$, and plot $B_{\rm p}$ versus $t$ of PSR J1634$-$4631 in Figure 4(a).

The decay rate $\dot{B}_{\rm p}$ of PSR J1634$-$4631 is also an important issue.  From Equation (20), we obtain the expression of $\dot{B}_{\rm p}$ and $t$ for the pulsar,
\begin{equation}
\frac{dB_{\rm p}(t)}{dt}=-\frac{B_{\rm p}(0)}{\tau_{\rm D}}e^{-\frac{t}{\tau_{\rm D}}}.
\end{equation}
The magnetic field decay rate $\dot{B}_{\rm p}$ of the pulsar declines with $t$, as shown in Figure 4(b). When $t=t_{\rm age}$, we get the present value of $dB_{\rm p}/dt\sim -(1.66-3.85)\times10^{8}$\,G\,yr$^{-1}$.
\begin{figure}[th]
\centering
 \vspace{0.6cm}
\subfigure[]{
    \label{evsb:a} 
    \includegraphics[width=7.2cm]{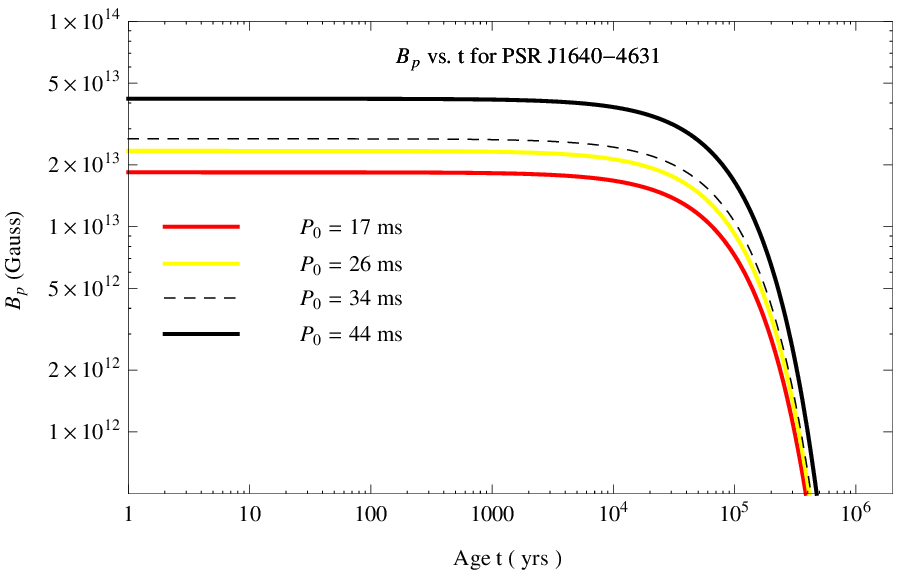}}
  \hspace{0.2mm}
  \subfigure[]{
    \label{evsb:b} 
    \includegraphics[width=7.2cm]{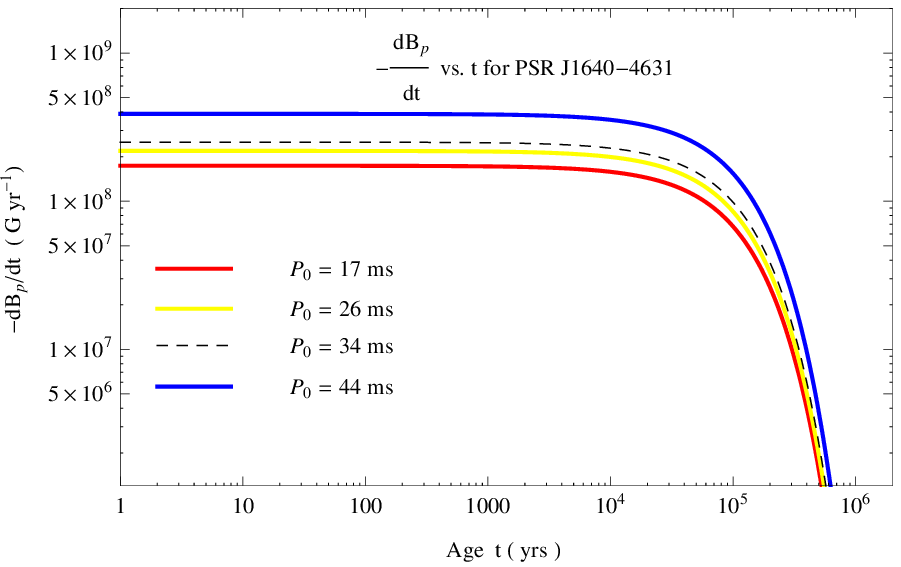}}
 \caption{(a)Relation between $B_{\rm p}$ and $t$ of PSR J1640$-$4631. (b)Relation between $\dot{B}_{\rm p}$ and $t$ of PSR J1640$-$4631.}
 \label{fig4}
 \end{figure}
\subsection{The relation of $\tau_{\rm c}$ and $t_{\rm age}$. }
The constraints on three parameters $B_{\rm p}(0)$, $t_{\rm age}$ and $\tau_{\rm D}$) of the pulsar also can be useful in accounting for the age difference between $\tau_{\rm c}$ and $t_{\rm age}$.
Inserting Equation (23) and Equation (24) into $\tau_{\rm c}=P/2\dot{P}$, we have
\begin{eqnarray}
\tau_{\rm c}=&&\frac{1}{2}(\frac{2\pi^{2}R^{6}B^{2}_{\rm p}(0)
}{3Ic^{3}})
\times e^{\frac{2t}{\tau_{\rm D}}}\times \nonumber\\
&&\left[P^{2}_{0}
+(\frac{2\pi^{2}R^{6}B^{2}_{\rm p}(0)
}{3Ic^{3}})\times\tau_{\rm D}\times
(1-e^{\frac{-2t}{\tau_{\rm D}}})\right].
\end{eqnarray}
If $\tau_{\rm D}\rightarrow \infty$, then Equation (28) is approximated as
\begin{equation}
\tau_{\rm c}=\frac{\pi^{2}R^{6}B^{2}_{\rm p}(0)
}{3Ic^{3}}\times\left[P^{2}_{0}
+(\frac{2\pi^{2}R^{6}B^{2}_{\rm p}(0)
}{3Ic^{3}})\times 2t\right].
\end{equation}
From Equations (28)-(29), we plot the diagrams of $\tau_{\rm c}$ vs. $t$ for the pulsar in Figure 5.
\begin{figure}[bt]
\centering
\includegraphics[angle=0,scale=.88]{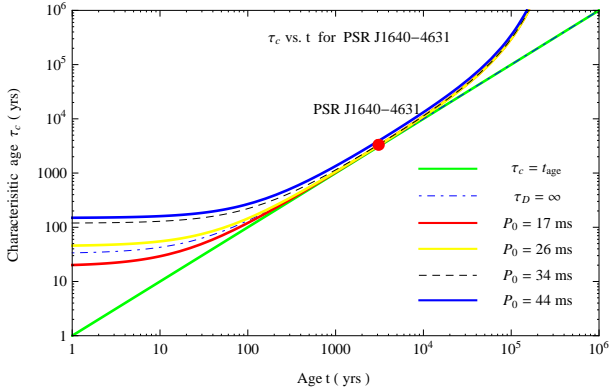}
\caption{Characteristic age as a function of $t$ for PSR J1640$-$4631. Here, the measured value of $\tau_{\rm c}$ is shown with the red dot. The error in data point is smaller than the size of the symbol. }
\label{tc}
\end{figure}
In Figure 5, the red dotted line denotes $\tau_{\rm c}=t_{\rm age}$.  It is easy to see that $\tau_{\rm c}$ increases with the age, but the fitted curves given by Equation (28) are always above the boundary of $\tau_{\rm c}=t_{\rm age}$. This is because the characteristic age of a pulsar is always higher than its inferred age if $n\geq 3$, which can be seen from Equation (5) in Section 2.
\subsection{Diagrams of $N-t$, $P-t$, $\dot{P}-t$ and $\ddot{P}-t$.}
Considering a pulsar with rotational angular momentum $L$\, ($L=I\Omega(t)$), the braking torque acting on the pulsar is given by
\begin{equation}
N=\frac{dL}{dt}=I\dot{\Omega(t)},
\end{equation}
where $I$ is assumed to be constant in time.
The decay of $B_{\rm p}$ inevitably causes a decrease in the dipole braking torque $N$, which is described by
\begin{eqnarray}
&&N=I\dot{\Omega}=-\frac{4\pi^{3}R^{6}B^{2}_{\rm p}(0)
}{3c^{3}}\times e^{\frac{-2t}{\tau_{\rm D}}}\times\nonumber\\
&&\left[P^{2}_{0}
+(\frac{2\pi^{2}R^{6}B^{2}_{\rm p}(0)}{3Ic^{3}})\times \tau_{\rm D}\times(1-e^{\frac{-2t}{\tau_{\rm D}}})\right]^{-3/2}.
\end{eqnarray}
If $\tau_{\rm D}\rightarrow \infty$, Equation (31) is rewritten as
\begin{equation}
N(t)=-\frac{4\pi^{3}R^{6}B^{2}_{\rm p}(0)
}{3c^{3}}\times
\left[P^{2}_{0}
+(\frac{2\pi^{2}R^{6}B^{2}_{\rm p}(0)
}{3Ic^{3}})\times 2t\right]^{-3/2}.
\end{equation}
Then we plot the diagrams of $N-t$ for PSR J1634$-$4631, as shown in Figure 6(a). It is obvious that $N$ decreases as $t$ increases.

As a comparison, below we give the relations of $P-t$, $\dot{P}-t$, and $\ddot{P}-t$ in the case of the MDR model:
\begin{equation}
P(t)=\left[P^{2}_{0}+(\frac{2\pi^{2}R^{6}B^{2}_{\rm p}(t)
}{3Ic^{3}})\times 2t \right]^{1/2},
\end{equation}
\begin{equation}
\dot{P}(t)=(\frac{2\pi^{2}R^{6}B^{2}_{\rm p}(t)
}{3Ic^{3}})
\times\left[P^{2}_{0}
+(\frac{2\pi^{2}R^{6}B^{2}_{\rm p}(0)
}{3Ic^{3}})\times 2t\right]^{-1/2}
\end{equation}
and
\begin{equation}
\ddot{P}(t)= -(\frac{2\pi^{2}R^{6}B^{2}_{\rm p}(0)
}{3Ic^{3}})^{2}\times \left[P^{2}_{0}
+(\frac{2\pi^{2}R^{6}B^{2}_{\rm p}(0)
}{3Ic^{3}})\times 2t\right]^{-3/2}.
\end{equation}
Based on Equations (22)-(24) and (31)-(33), we numerically simulate the relations of $P-t$, $\dot{P}-t$ and $\ddot{P}-t$ for PSR J1634$-$4631.
\begin{figure*}[th] %
\begin{center}
\begin{tabular}{cc}
\scalebox{0.76}{\includegraphics{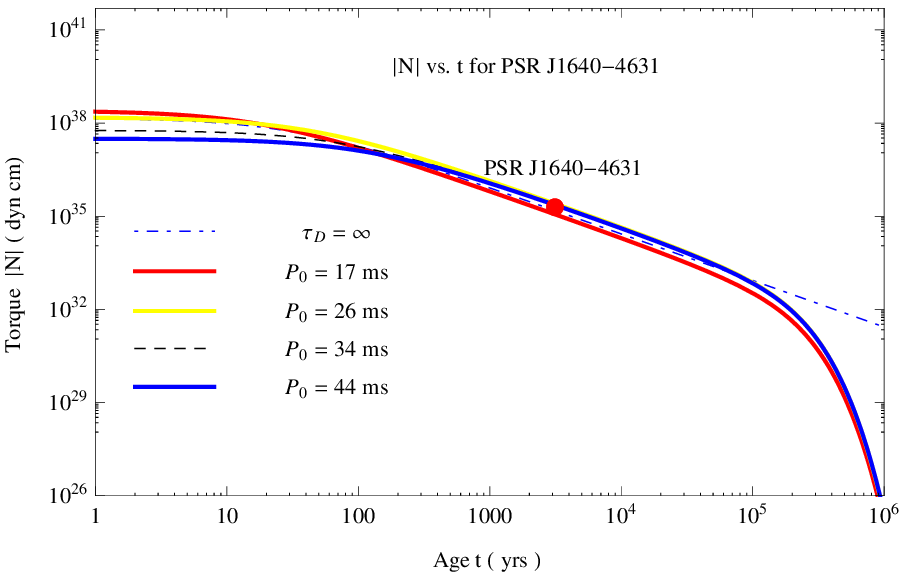}}&\scalebox{0.76}{\includegraphics{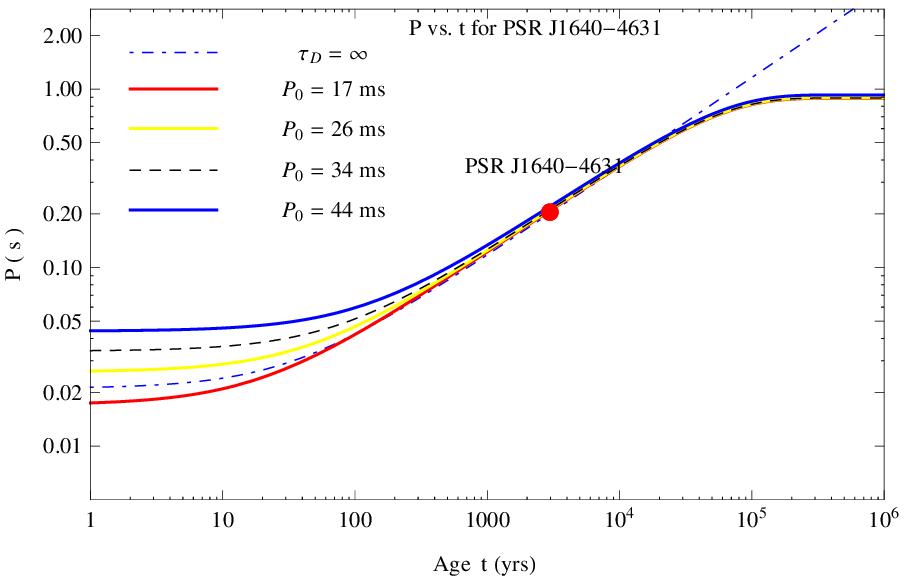}}\\
(a)&(b)\\
\scalebox{0.76}{\includegraphics{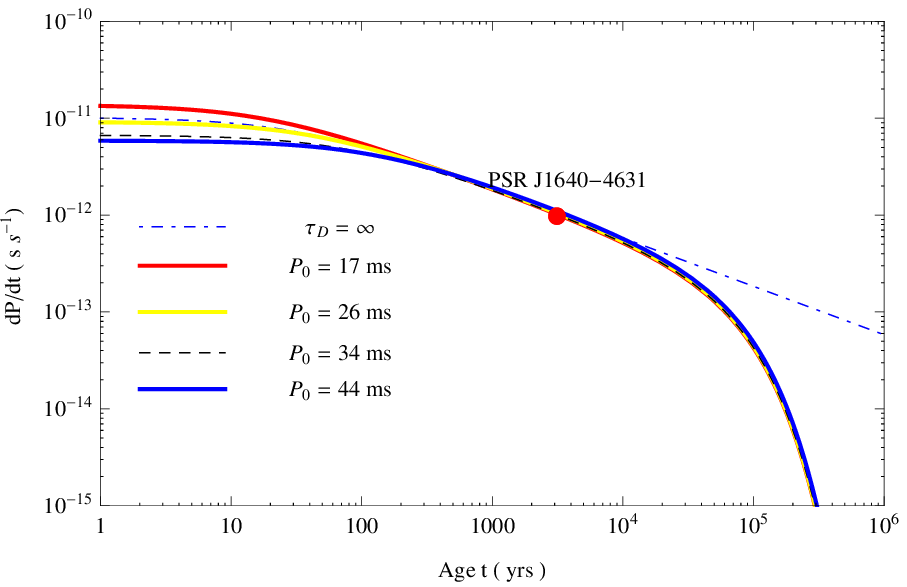}}&\scalebox{0.76}{\includegraphics{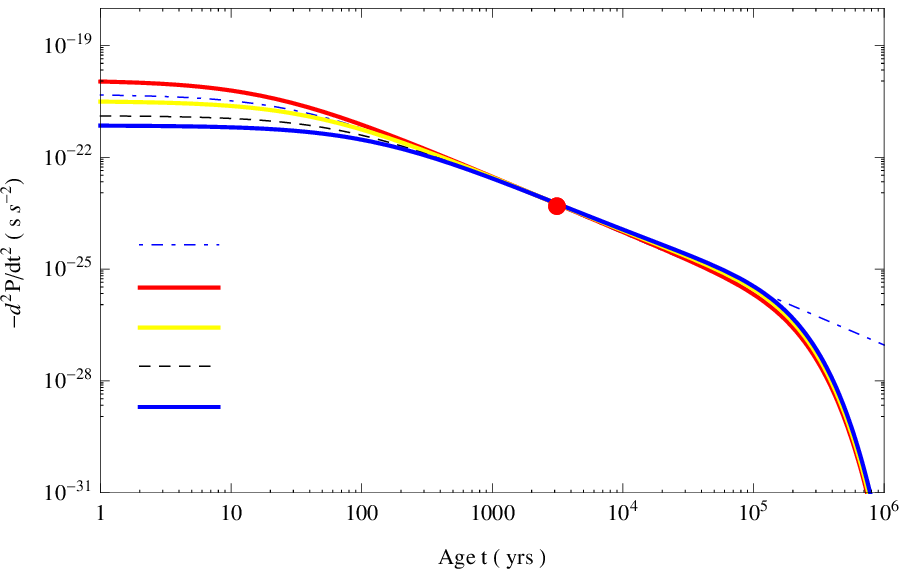}}\\
(c)&(d)\\
\end{tabular}
\end{center}
\caption{Numerical simulations of $N-t$, $P-t$, $\dot{P}-t$ and $\ddot{P}-t$ for PSR J1634$-$4631. Panels(a), (b), (c), (d) describe the relations of $N-t$ $P-t$, $\dot{P}-t$ and $\ddot{P}-t$ for the pulsar, respectively.
 \label{5:fig}}
\end{figure*}
In Figure 6, the measured values of $N$, $P$, $\dot{P}$ and $\ddot{P}$ are shown with the red dots, and the errors in data point are smaller than the sizes of the symbols. In the terms of the dipole magnetic field decay model, as $t$ increases, $P$ increases rapidly at the earlier evolution stage and then increases slowly at the later evolution stage, whereas $\dot{P}$ and $\ddot{P}$ decrease slowly at the earlier evolution stage and  then decrease rapidly at the later evolution stage. Unlike PSR J1734$-$3333, which could evolve into a magnetar (Espinoza et al. 2011), PSR J1634$-$4631 will eventually come down to the death valley, due to the pulsar death effect.
\section{Alternative models for the dipole magnetic field decay}
It is worth emphasizing that our results and conclusions strongly depend on Equation (20) in which an exponential form of dipole magnetic field  decay is assumed. In order to investigate how different magnetic field decay laws would affect our results, we assume that the dipole magnetic is deeply buried in the inner crust and extends into the core (see, e.g., Muslimov \& Page 1995; Gourgouliatos \& Cumming 2015). If the magnetic field decays via the Hall drift and ohmic decay with a nonlinear form:
\begin{equation}
 B_{\rm p}(t) = \frac{B_0}{1+\,t/\tau_{\rm D}}.
 \end{equation}
As a comparison, we give the relations of $P-t$, $\dot{P}-t$ and $\ddot{P}-t$ in this nonlinear decay form. Inserting Equation (36) into Equation (22), we obtain the expression of the spin period,
 \begin{equation}
P(t)=\left[P^{2}_{0}+(\frac{4\pi^{2}R^{6}B^{2}_{\rm p}(0)}{3Ic^{3}})\times
\frac{\tau_{\rm D}\cdot t}{\tau_{\rm D}+t}\right]^{1/2}.
\end{equation}
Utilizing the differential method, we get the first-order
derivative of the spin period $\dot{P}(t)$,
\begin{eqnarray}
&&\dot{P}(t)=(\frac{2\pi^{2}R^{6}B^{2}_{\rm p}(0)
}{3Ic^{3}})\times (\frac{\tau_{\rm D}}{t+\tau_{\rm D}})^{2} \nonumber\\
&&\times \left[P^{2}_{0}
+(\frac{4\pi^{2}R^{6}B^{2}_{\rm p}(0)
}{3Ic^{3}})\times \frac{\tau_{\rm D}\cdot t}{\tau_{\rm D}+t}\right]^{-1/2},
\end{eqnarray}
and the second-order derivative of the spin period $\ddot{P}(t)$,
\begin{eqnarray}
&&\ddot{P}(t)=-(\frac{2\pi^{2}R^{6}B^{2}_{\rm p}(0)
}{3Ic^{3}})^{2}\cdot(\frac{\tau_{\rm D}}{t+\tau_{\rm D}})^{4}\cdot[P^{2}_{0}+ \nonumber\\
&& (\frac{4\pi^{2}R^{6}B^{2}_{\rm p}(0)}{3Ic^{3}})\cdot
\frac{\tau_{\rm D}\cdot t}{\tau_{\rm D}+t}]^{-3/2}-(\frac{4\pi^{2}R^{6}B^{2}_{\rm p}(0)
}{3Ic^{3}})\cdot\tau_{\rm D}^{2}\nonumber\\
&&\times(\frac{1}{t+\tau_{\rm D}})^{3}
\cdot[P^{2}_{0}+(\frac{4\pi^{2}R^{6}B^{2}_{\rm p}(0)}{3Ic^{3}})\cdot
\frac{\tau_{\rm D}\cdot t}{\tau_{\rm D}+t}]^{-1/2}.
\end{eqnarray}
If we assume $M\sim$ 1.0-2.2$\,M_{\bigodot}$ and use the same APR3 EOS, then we obtain the effective magnetic field decay timescale $\tau_{\rm D}\sim0.91(8)\times10^{5}$\,yr and the true age $t_{\rm age}\sim 3100-3200$\,yr, the initial dipole magnetic field $B_{\rm p}(0)\sim(1.72-3.83)\times 10^{13}$\,G, and the dipole magnetic field decay rate $dB_{\rm p}/dt\sim -(1.76-3.94)\times10^{8})$\,G\,yr$^{-1}$.

Alternatively, the magnetic field is assumed to decay with a power-law decay form:
\begin{equation}
 B_{\rm p}(t)= B_{\rm p}(0)\times(t/\tau_{\rm D})^{\varepsilon},
 \end{equation}
 as done in Muslimov \& Page (1995,\,1996), where $\varepsilon<0$ is the magnetic field index. In this case, Equations (37)-(39) are replaced by
 \begin{equation}
P(t)=\left[P^{2}_{0}+(\frac{4\pi^{2}R^{6}B^{2}_{\rm p}(0)}{3Ic^{3}})\times
\frac{\tau_{\rm D}}{2\varepsilon+1}\times(\frac{t}{\tau_{\rm D}})^{2\varepsilon+1}\right]^{1/2},
\end{equation}
\begin{eqnarray}
&&\dot{P}(t)=(\frac{2\pi^{2}R^{6}B^{2}_{\rm p}(0)
}{3Ic^{3}})\times(\frac{t}{\tau_{\rm D}})^{2\varepsilon}\times \nonumber\\
&& \left[P^{2}_{0}+(\frac{4\pi^{2}R^{6}B^{2}_{\rm p}(0)}{3Ic^{3}})\times
\frac{\tau_{\rm D}}{2\varepsilon+1}\times(\frac{t}{\tau_{\rm D}})^{2\varepsilon+1}\right]^{-1/2},
\end{eqnarray}
and
\begin{eqnarray}
&&\ddot{P}(t)=-(\frac{2\pi^{2}R^{6}B^{2}_{\rm p}(0)
}{3Ic^{3}})^{2}\times \frac{2\varepsilon}{\tau_{\rm D}}\times(\frac{t}{\tau_{\rm D}})^{2\varepsilon-1}\times  \nonumber\\
&& \left[P^{2}_{0}+(\frac{4\pi^{2}R^{6}B^{2}_{\rm p}(0)}{3Ic^{3}})\times
\frac{\tau_{\rm D}}{2\varepsilon+1}\times(\frac{t}{\tau_{\rm D}})^{2\varepsilon+1}\right]^{-1/2}\nonumber\\
&&-(\frac{2\pi^{2}R^{6}B^{2}_{\rm p}(0)
}{3Ic^{3}})^{2}\times (\frac{t}{\tau_{\rm D}})^{4\varepsilon}\times \nonumber\\
&&\left[P^{2}_{0}+(\frac{4\pi^{2}R^{6}B^{2}_{\rm p}(0)}{3Ic^{3}})\times
\frac{\tau_{\rm D}}{2\varepsilon+1}\cdot(\frac{t}{\tau_{\rm D}})^{4\varepsilon+1}\right]^{-3/2},
\end{eqnarray}
 respectively.  For the sake of simplicity, we assume that the range of $t_{\rm age}$ obtained from the power-law decay form (Equation (40)) is very close to those obtained from the exponential and nonlinear decay forms (Equation (20) and Equation (36)), e.g., $t_{\rm age}\sim 2800-3200$\,yr. Utilizing the same NS mass range and fitting method, we get the magnetic field index $\varepsilon=-0.034(1)$, the effective field decay timescale $\tau_{\rm D}\sim (0.4-2.5)\times 10^{6}$\,yr, the initial dipole magnetic field strength $B_{\rm p}(0)\sim (1.44-3.03)\times 10^{13}$\,G, and the magnetic field decay rate $dB_{\rm p}/dt\sim -(1.78-4.42)\times10^{8})$\,G\,yr$^{-1}$.

It is worth addressing that for ordinary radio pulsars the effect of field decay may not be very pronounced, and so it is difficult to uncover it. For example, Mukherjee \& Kembhavi (1997) synthesized a population of pulsars using assumed theoretical distributions of age, initial magnetic field, period of rotation, position, luminosity, and dispersion measure and concluded that the timescale for the field decay must be greater than 160 Myr. Other studies (Bhattacharya et al. 1992; Hartman et al. 1997; Regimbau \& de Freitas Pacheco 2001) also confirm these long timescales. We recall that the pulsars considered by these authors are older radio pulsars having the kinetic ages and/or characteristic ages $\tau\sim 10-100$\,Myr, the dipole magnetic field strengthes $B\sim (2-5)\times10^{12}$\,G, which are different from young X$-$ray/$\gamma-$ray pulsar PSR J1640$-$4631 with $\tau_{\rm c}\sim 3360$\,yr\,(the true age $t_{\rm age}\sim 2800-3200$\,yr). Furthermore, using the selection effect of pulsars' distance distribution and velocity distribution, Mukherjee \& Kembhavi (1997) obtained the data for radio pulsars from the catalog of pulsars in the Princeton database (Taylor et al.\,1995) and dropped binaries, X-ray pulsars and gamma-ray pulsars from the simulated pulsar samples. Thus, the  magnetic field decay timescale obtained from Mukherjee \& Kembhavi (1997) is larger than the one we give. Nevertheless, the magnetic field decay timescale of pulsars is an interesting subject worth long-term study.
\section{Comparisons and Conclusions }
In this work, the high braking index of PSR J1640$-$4631 is well interpreted within a combination of MDR and dipole magnetic field decay. The related comparisons and conclusions are given as follows.
\subsection{Comparisons with our previous work  \label{subsec:a}}
In our previous work (Gao et al. 2016), we investigated the braking indices of eight magnetars. We attributed the larger braking indice\,($n>3$) of three magnetars to the decay of external braking torque, which might be caused by magnetic field decay, magnetospheric current decay, or the decay of magnetic inclination angle, and also calculated the dipolar field decay rates, which are compatible with the measured values of $n$ for the following three magnetars:
\begin{table}
\caption{ Dipole Field Decay Rates of the  Magnetars.  }
\begin{tabular}{ccc}\\
\hline 
 Source & Braking index $n$  &$\frac{dB_{\rm p}}{dt}$\,(G\,yr$^{-1})$\\
 \hline
1E 1841$-$045 & $9-17$  & $-(6.55\times10^{11}-1.32\times10^{12})$   \\
SGR 0501$+$4516 & $4.6-8.0$ &$-(5.19\times10^{9}-9.66\times10^{10}$ \\
1E 2259$+$586 & $22-42$ &$-(2.44\times10^{9}-4.76\times10^{9})$ \\
\hline\\
\end{tabular}\\
\label{tb:2}
\end{table}
Compared with PSR J1640$-$4631, the three magnetars in Table 2 have higher magnetic field decay rates and larger braking indices. The main reasons are twofold:(i)\,magnetars are a kind of special pulsar powered by their magnetic energy rather than their rotational energy (Duncan \& Thompson 1992; Gao et al. 2013; Liu 2017); and (ii)\,to account for the high-value braking indices of magnetars, we adopted the updated magnetothermal evolution model of Vigan\`{o} et al. (2013) (see Gao et al. 2016 for details).

In addition, we show the long-term rotational evolution of PSR J1634$-$4631 in Figure 7. Five fitted lines in Figure 7 represent the evolution paths of the pulsar in the context of dipole magnetic field decay, and the arrows represent the spin-down evolution directions with different initial spin period in the next 100 kyr.
PSR J1634$-$4631 would be placed in the top left region of the $P-\dot{P}$ diagram at its birth. As it spins down, the pulsar moves toward the bottom right of the $P-\dot{P}$ diagram in $\sim10^5-10^6$\,yr and will eventually come down to the death valley, based on the dipole magnetic field decay model.

\begin{figure}[bt]
\centering
\includegraphics[angle=0,scale=.88]{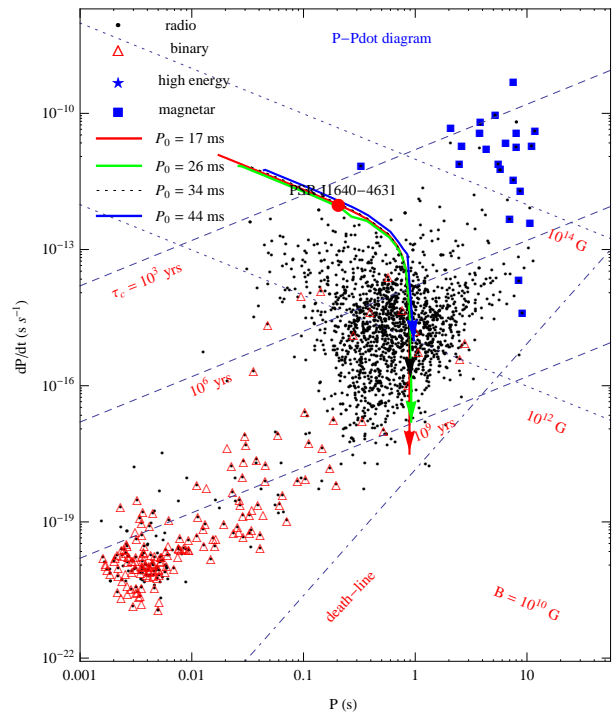}
\caption{Long-term rotational evolution of PSR J1640$-$4631 dominated by the dipole magnetic field decay. Radio, binary and magnetars are defined by black dot, red triangle, and blue square, respectively. The red solid line, green solid line, black dotted line, and blue solid line
stand for the predictions of the dipole magnetic field decay
model with $P_{0}=17, 26, 34$ and 44 ms, respectively.
The red filled circle denotes the observations of PSR J1640$-$4631.}
\label{P-Pdot}
\end{figure}
\subsection{Comparsions with other models  \label{subsec:b}}
Recently, to understand better the existence of both pulsar braking indices both larger than 3 and smaller than 3, different braking mechanisms have been proposed (e.g., Ek\c{s}i et al. 2016; Chen 2016; Clark et al. 2016; Coelho et al. 2016; de Araujo et al. 2016 a, b; Magalhaes et al. 2016; Tong \& Kou 2017). For the purpose of illustrating the feasibility of this theoretical model, it is worth comparing our results with those of other models.
\begin{enumerate}
\item Initial spin period estimate.
 Recently, by numerically simulating, Gotthelf et al. (2014) predicted that PSR J1640$-$4631 has a short initial spin period $P_{0}\sim 15$\,ms. Such a short $P_{0}$ requires a smaller braking index ($n<3$) and a higher true age ($t_{\rm age}>\tau_{\rm c}$)\,(Gotthelf et al. 2014), which departures from the observed $n$ and $\tau_{\rm c}$ of the pulsar obviously.
 In this paper, we firstly introduce a mean braking index $\overline{n}$ and a mean rotation energy conversion coefficient $\overline{\zeta}$ and then estimate the initial spin period $P_{0}=17-44$\,ms. The largest uncertainty of our model comes from $\overline{\zeta}$, because the true value of $\overline{\zeta}$ is unable to be obtained. However, if $P_{0}\ll P$, the uncertainty of $\overline{\zeta}$ will have little effect on the results of the three parameter of $t_{\rm age}$, $\tau_{\rm D}$ and $B_{\rm p}(0)$, according to our model. In this work, the initial spin period ($P_{0}\sim 17-44$\,ms) is obtained by combining pulsar's high- energy observations with the EOS.
\item True age estimate.  Recently, using the leptonic
PWN model\footnote{Slane et al. (2010) modeled the $\gamma-$ray emission from HESS J1640$-$465 and the associated broad-band spectrum with an evolving, one-zone PWN model, where the $\gamma-$rays are ambient photons inverse Compton scattered by relativistic electrons that also produce the synchrotron emission.}, Slane et al. (2010) assumed the age of G338.3$-$0.0 $t_{\rm SNR}\sim$10\,kyr, and Gotthelf et al. (2014) gave an age $t_{\rm SNR}\sim 6.8$\,kyr. The former requires a constant braking index $n=3$, while the latter requires a smaller braking index $n\approx 2.0$\,(Gotthelf et al.2014).  However, their methods are universally based on supernova explosion mechanisms, due to the unknown explosion energy $E$ and the ambient mass density $\rho_{0}$. In this work, the true age estimate ($t_{\rm age}\sim 2900-3100$\,yr) is obtained by solving the pulsar spin-down evolution equation set.
\item Inclination angle estimate.  Based on the MDR model with corotating plasma, Ek\c{s}i et al. (2016) found that the high braking index is consistent with two different inclination angles, $\alpha\sim 18.5(3)^{\circ}$ and $56(4)^{\circ}$. Very recently, based on the vacuum gap model and low mass ($M\sim 0.1\,M_{\bigodot}$) neutron star candidate, employing the vacuum gap model, Chen (2016) proposed that a low-mass neutron star\,($M\sim 0.1M_{\bigodot}$) with a large inclination angle (close to the perpendicular case) could interpret the high braking index, as well as the radio-quiet nature of the pulsar. Note that in these two models a constant magnetic field has been uniformly assumed for PSR J1640$-$4631, and the discrepancy of $\alpha$ they obtained is obvious. As we know, it is the most meaningful to determine a pulsar's inclination angle from its timing and polarization observations. We expect that PSR J1640$-$4631 has a reliable measurement of $\alpha$ from the future timing and polarization observations.
\end{enumerate}
\subsection{No detected bursts in PSR J1640$-$4631\label{subsec:c}}
Since the higher-order magnetic moments decay rapidly with the distance to the stellar center, the influence of multipole magnetic moments of the star on the braking index can be ignored (if multipole magnetic fields exist in the pulsar). Interestingly, Chen (2016) supposed that there could be superhigh multipole magnetic fields inside PSR J1640$-$4631 (e.g., quadrupole magnetic field) resulting in a large stellar deformation and strong GW emission. If this is true, the burst behaviors should have been detected. However, up to date, no burst has been observed. If our model is correct, the undetected bursts in PSR J1640$-$4631 will be well explained as follows: by using the currently estimated values of $B_{\rm p}$ and $\dot{B}_{\rm p}$, we estimate the magnetic field energy decay rate, $dE_{B}/dt=\frac{d}{dt}(\frac{B^{2}}{8\pi})\times\frac{4}{3}\pi R^{3}=(1.2-2.8)\times10^{32}$\,erg\,s$^{-1}$, where $B^{2}_{\rm p}/8\pi$ is the magnetic field energy density. This magnetic field energy decay rate is about two orders of magnitude lower than the X-ray luminosity of PSR J1640$-$4631. Nevertheless, our prediction could be tested by the future high-energy observations of the pulsar.

 In this work, we interpret the high braking index of PSR J1640$-$4631 with a combination of the dipole magnetic field decay and MDR models, ignoring the influences of all other possible braking torques. It is suggested that the decay of dipole magnetic field could occur universally in pulsars with $n<3$. If so, why are the measured braking indices of the eight pulsars lower than 3 (see, e.g., Gao et al. 2016; Tong \& Kou 2017)? Below we suggest two possible reasons.
 \begin{enumerate}
\item The outflowing particle winds (the relativistic outflow mainly composed of the electron-positron pairs) luminosities in the eight rotationally powered pulsars with low braking indices may be so high that additional braking torques provided by winds can be comparable to the magneto-dipole braking torques in magnitude. Thus the stars' braking indices are less than 3. A PWN is dominated by a bright collimated feature, which is interpreted as a relativistic jet directed along the pulsar spin axis\,(e.g., Gaensler et al. 2003; Grondin et al. 2013). The PWNs of five low braking index pulsars have been detected.\footnote{PSR B0531$+$21\,(Crab) with $n=2.34(2)$, Lyne et al. 1993, 2015; PSR B0833$-$45\,(Vela) with $n=1.4(2)$, Lyne et al. 1996; PSR B1509$-$58 with $n=2.839(1)$, Livingstone et al. 2007; PSR J1833$-$1034 with $n=1.857(6)$, Roy et al. 2012, and PSR B0540$-$69 with $n=2.140(9)$, Ferdman et al. 2015} Especially, the spin-down luminosity of the Crab pulsar $\sim 5\times 10^{38}$\, erg\,s$^{-1}$ is converted into radiation with a remarkable efficiency, approaching $\sim30\%$\,(Abdo et al. 2011). No detection of PWNs for three pulsars (PSR J1846$-$0258 with  $n=2.65(1)$, Livingstone et al. 2007; PSR J1119-6127 with  $n=2.684(2)$, Weltevrede et al. 2011; and PSR J1734$-$3333 with $n=0.9(2)$, Espinoza et al. 2011) could be explained by the tenuous interstellar medium around the pulsars.

\item Another possibility is that the eight pulsars are experiencing toroidal magnetic field\,(e.g., octupole field) decay via Hall drift and ohmic dissipation. For example, magnetar-like outbursts from three pulsar of PSR J1846$-$0258, PSR J1734$-$3333 and PSR J1119-6127 were reported\,(e.g., Gavriil et al. 2008; G\"{o}\v{g}\"{u}\c{s} et al.2016). These outbursts could be caused by the decay of initial toroidal multipole magnetic fields. The multipole magnetic fields are merging through crustal tectonics to form a dipole magnetic field causing a low braking index.
\item For PSR J1640$-$4631, though it has an associated PWN with very low rotational energy transformation coefficient\,(see Section 3 of this work), the pulsar may be at an inactive wind epoch. Maybe this pulsar has experienced the stage of multipole fields merging, i.e., its dipole magnetic field stops increasing. At the current epoch, the emission and braking properties of the pulsar are dominated by the dipole magnetic field decaying, which causes a high mean braking index. In summary, attempts to explain the lower braking indices\,($n< 3$) of pulsars within the dipole magnetic field decay model coupled with wind braking and/or toroidal field decaying will be considered in our future studies.
\end{enumerate}
\subsection{Conclusions\label{subsec:d}}
In this work, we interpreted the high braking index of PSR J1640$-$4631 with a combination of the MDR and dipole magnetic field decay. By introducing a mean rotation energy conversion coefficient $\overline{\zeta}$ and combining the EOS with the high-energy and timing observations of the pulsar, we give the constrained values of $P_{0}$, $t_{\rm age}$, $\tau_{\rm D}$ and $B_{\rm p}(0)$, and numerically simulat the dipole magnetic field and spin-down evolutions of the pulsar. The high braking index of $3.15(3)$ of PSR J1640$-$4631 is attributed to its long-term dipole magnetic field decay at a low rate $dB_{\rm p}/dt\sim -(1.66-3.85)\times10^{8}$ G yr$^{-1}$. Considering the uncertainties and assumptions in this work, thus our theoretical model needs to be tested and modified by the future polarization, timing, and high-energy observations of PSR J1640$-$4631. Our results may apply to other pulsars with higher braking indices.
\acknowledgements
We thank an anonymous referee for carefully reading
the manuscript and providing valuable comments that improved this paper substantially. We also thank Prof. Andrew Lyne for useful discussions. This work was supported by National Basic Research Program of China grants 973 Programs 2015CB857100; the West Light
Foundation of CAS through grants XBBS-2014-23, XBBS-2014-22 and 2172201302; Chinese National Science Foundation through grants No.11673056,11622326, 11273051, 11373006, 11133004 and 11173042; the Strategic Priority Research Program of CAS through no.XDB23000000 and National Program on Key Research and Development Project through no. 2016YFA0400803.


\label{lastpage}
\end {document}